\documentclass[preprint,12pt]{elsarticle}

\usepackage{graphicx}
\usepackage{amssymb}

\journal{Icarus}

\begin{document}

\begin{frontmatter}


\title{Cometary ions detected by the Cassini spacecraft 6.5 au downstream of Comet 153P/Ikeya-Zhang}



\ead{g.h.jones@ucl.ac.uk}

\author[lab1,lab2]{G. H. Jones}
\author[lab3]{H. A. Elliott}
\author[lab4]{D. J. McComas}
\author[lab8]{M. E. Hill}
\author[lab8]{J. Vandegriff}
 \author[lab6]{E. J. Smith\fnref{label2}}
\author[lab7]{F. J. Crary}
\author[lab3]{J. H. Waite}

 \fntext[label2]{Deceased}

\address[lab1]{Mullard Space Science Laboratory, Department of Space \& Climate Physics, University College London, Holmbury St. Mary, Dorking, Surrey RH5 6NT, UK}
\address[lab2]{The Centre for Planetary Sciences at UCL/Birkbeck, London, UK}
\address[lab3]{Southwest Research Institute, San Antonio, TX, USA}
\address[lab4]{Department of Astrophysical Sciences, Princeton University, Princeton, NJ 08544, USA}
\address[lab6]{Jet Propulsion Laboratory, California Institute of Technology, Pasadena, CA, USA}
\address[lab7]{Laboratory for Atmospheric and Space Physics, University of Colorado Boulder, CO, USA}
\address[lab8]{Johns Hopkins Applied Physics Laboratory, Laurel, MD, USA}

\begin{abstract}
During March-April 2002, while between the orbits of Jupiter and Saturn, the Cassini spacecraft detected a significant enhancement in pickup proton flux. The most likely explanation for this enhancement was the addition of protons to the solar wind by the ionization of neutral hydrogen in the corona of comet 153P/Ikeya-Zhang. This comet passed relatively close to the Sun-Cassini line during that period, allowing pickup ions to be carried to Cassini by the solar wind. This pickup proton flux could have been further modulated by the passage of the interplanetary counterparts of coronal mass ejections past the comet and spacecraft. The radial distance of 6.5 Astronomical Units (au) traveled by the pickup protons, and the implied total tail length of $>$7.5 au make this cometary ion tail the longest yet measured.

\textbf{Highlights}

\begin{itemize}
\item Pickup protons detected in the solar wind by the Cassini spacecraft at 7 au from the sun
\item Cassini was 6.6 au downstream of where Comet 153P/Ikeya-Zhang had passed
\item Geometry and solar wind speeds consistent with protons originating at the comet's hydrogen corona
\end{itemize}

\end{abstract}




\end{frontmatter}


\section{Introduction}

	The most extensive component of a comet's neutral coma is its hydrogen corona, which consists of fast atomic hydrogen released by the photodissociation of cometary water and its products, e.g. \citet{combi2008} and references therein. The atoms in this hydrogen cloud are eventually ionized by photo- and electron impact-ionization. The resultant protons, H$^{+}$, are picked up by the solar wind and carried antisunward, forming an extended but invisible H$^{+}$ tail that envelops the comet's main ion tail.  
   
\begin{figure}[h]
\centering\includegraphics[width=1.0\linewidth]{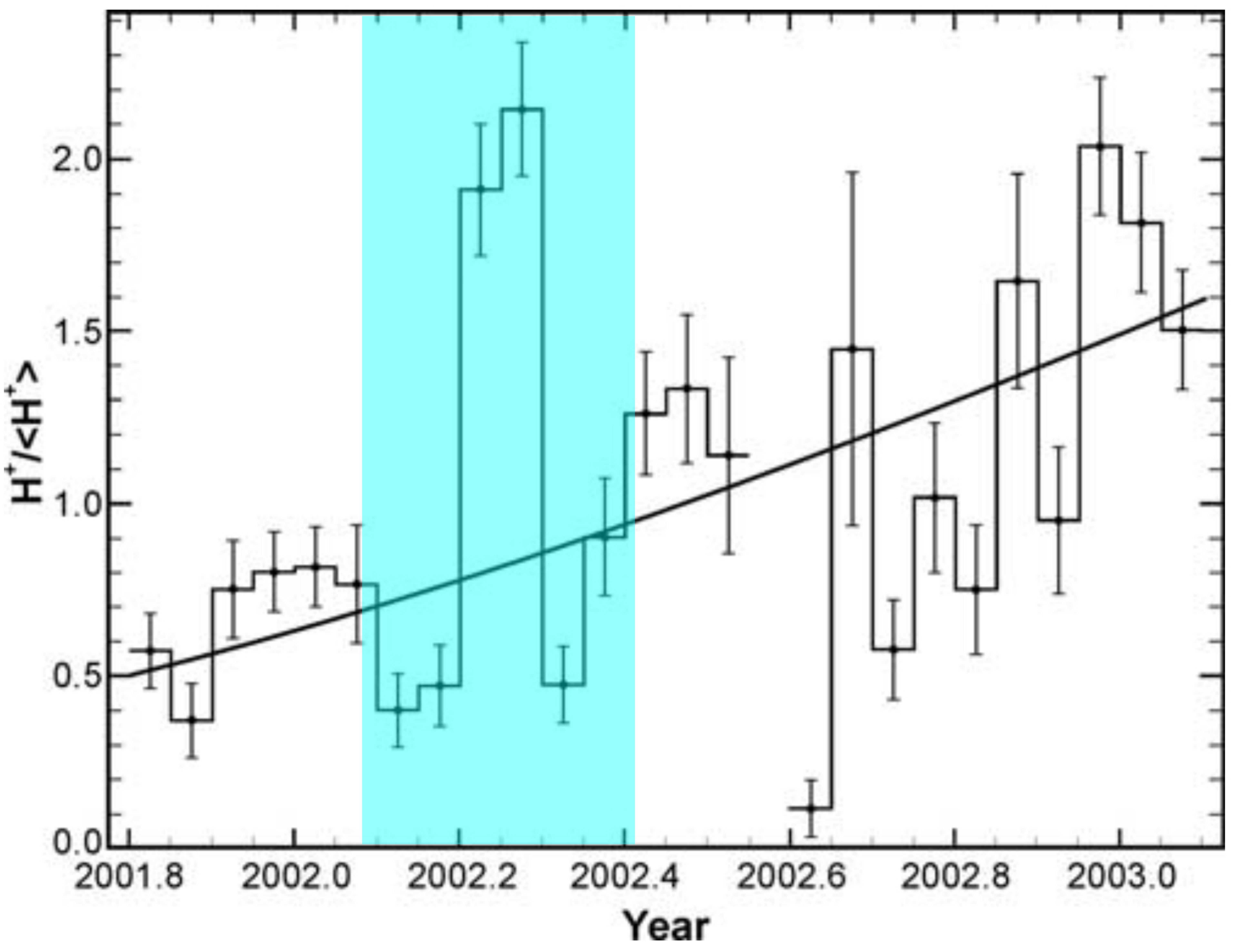}
\caption{Measured flux of pickup H$^{+}$ from late 2001 to early 2003, divided by its average value, first presented as Figure 5 of \citet{mccomas2004}. The missing data point is for a time when no unambiguous pickup H$^{+}$ data were available. The observed and modeled (smooth curve, see \citep{mccomas2004}) values show a factor of 2 increase as Cassini moved away from the center of the interstellar hydrogen shadow. The shaded region highlights the time period encompassing a statistically-significant increase in detected pickup protons, corresponding to the entire period covered in Figure~\ref{fig:fig1}.}
 \label{fig:newfig1}
\end{figure}

    During the Cassini spacecraft's interplanetary cruise to Saturn, several of its instruments obtained solar wind measurements. From October 2001 to February 2003, between 6.4 and 8.3~au from the Sun, the Cassini Plasma Spectrometer, CAPS \citep{young2004}, operated in a mode optimized for the detection of pickup H$^{+}$ and He$^{+}$. As the CAPS ion spectrometer aperture was rarely oriented toward the direction of arrival of the solar wind during this period, few solar wind ion measurements were available from this instrument. However, pickup ions were often detectable due to their more isotropic distributions. During the observation period, Cassini was leaving the helium focusing cone and hydrogen shadow that lie opposite the inflow direction of interstellar gas into the solar system. During this time, CAPS recorded an increase in pickup H$^{+}$ and decrease in He$^{+}$, constituting the first mass-resolved observations of interstellar pickup ions beyond the orbit of Jupiter \citep{mccomas2004}. The changes in detection rates of H$^{+}$ and He$^{+}$ during this period were not monotonic. In particular, a notable and statistically-significant increase in H$^{+}$ detections occurred during approximately 2002.2 -- 2002.3 ($\sim$March 14 -- April 19, or day of year, DOY 73 - 109), shown in Figure~\ref{fig:newfig1}.

This increase appears too large to be part of the general enhancement in H$^{+}$ flux expected as the hydrogen shadow was exited by Cassini. Possible causes of this large deviation were investigated. 
We discovered that in early March 2002, comet 153P/Ikeya-Zhang had passed upstream of the location that Cassini would occupy a few weeks later during the time of enhanced H$^{+}$ detections. This comet had been very active during that period, with a peak water production rate, Q(H$_{2}$O), of $\sim$10$^{30}$~molec~s$^{-1}$ at 0.51~au from the Sun. Here, we present evidence that the H$^{+}$ enhancement was due to cometary pickup protons from 153P joining the interstellar proton component and being detected by Cassini at a distance downstream of the nucleus of 6.55 au, or 9.8$\times$10$^{8}$~km.

\section{Observations}

\begin{figure}[h]
\centering\includegraphics[width=0.9\linewidth]{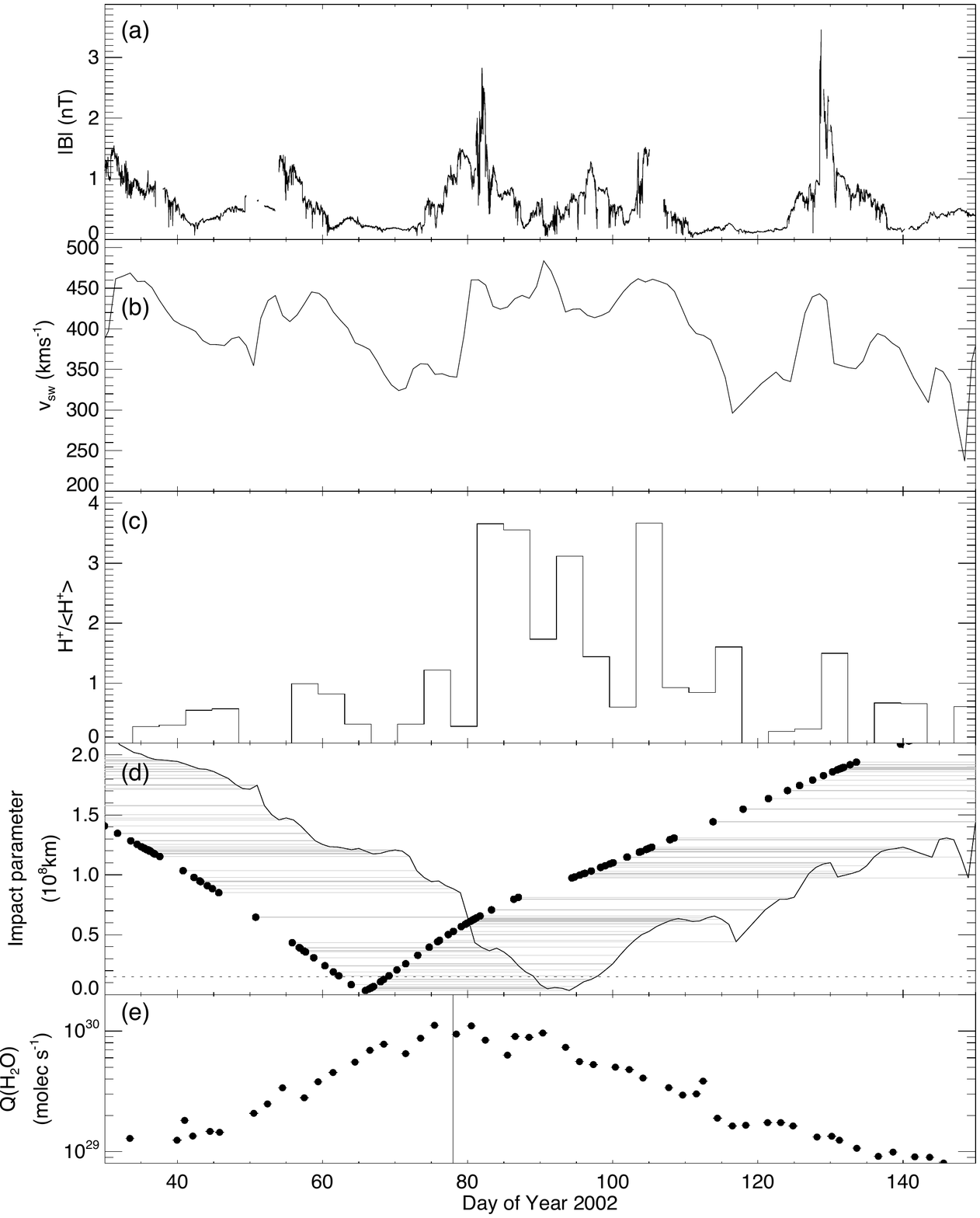}
\caption{Observations made at Cassini, the degrees of alignment between spacecraft and comet, and activity of comet 153P during early 2002. (a) Magnitude of the heliospheric magnetic field at Cassini. (b) Solar wind speed at Cassini measured by the MIMI instrument.
(c) Rate of pickup proton detection at Cassini obtained by the CAPS instrument, plotted with respect to the mean rate during the sampling period. A significant enhancement in H$^{+}$ fluxes is apparent shortly after DOY 80. (d) The closest distance that the solar wind plotted in the preceding panel (c) would have approached to 153P {\it en route} to Cassini (solid line), plotted against time of arrival at the spacecraft. Each distance data point is connected by a horizontal line to a solid dot showing the estimated closest approach time to the comet, assuming a purely radial flow. (e) Independently-determined water production rate of 153P from SOHO SWAN observations of the Lyman $\alpha$ corona, taken from \citet{combi2008}. 1$\sigma$ errors on the rates are smaller than the sizes of the plot symbols. The vertical line marks the time of perihelion of the comet.}
 \label{fig:fig1}
\end{figure}

Figure~\ref{fig:fig1} shows the rate of pickup H$^{+}$ detection at Cassini during four months of 2002, during which Cassini was between 6.9 and 7.3 au from the Sun. The CAPS instrument's response at its highest energies was not sufficiently well determined for this period to provide absolute values for the pickup H$^{+}$ flux. The latter is therefore shown relative to the mean flux during October 2001 to February 2003. Bins 3.65 days wide are used due to the low pickup proton detection rate. The H$^{+}$ flux clearly shows a relatively abrupt increase during March 22 - 25, DOY 81.30 -- 84.95, with rates remaining generally elevated for approximately 30 days after the initial increase. Also plotted is the magnetic field magnitude measured by the spacecraft's magnetometer \citep{dougherty2004}. We discuss the apparent correlations between changes in the latter and the proton flux later. 

\begin{figure}[h]
\centering\includegraphics[width=1.0\linewidth]{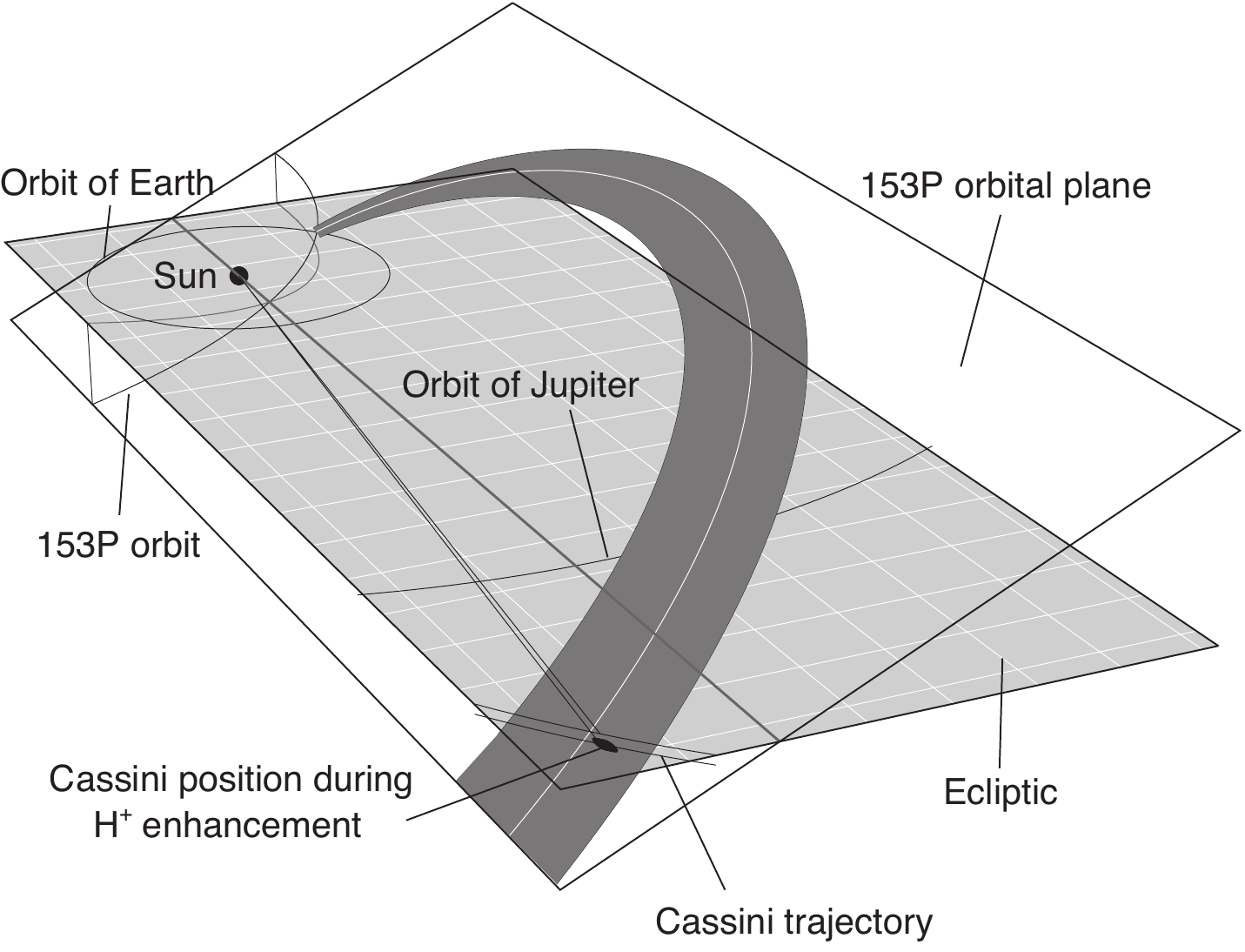}
\caption{Geometry of the near-alignment of comet and spacecraft. Shown are the relative positions of the comet's orbit and Cassini, together with a representative proton tail profile, for 2002 April 2 with a constant solar wind velocity of 400 kms$^{-1}$. A 0.1~au tail width at the comet's head was assumed, with radial expansion of the solar wind with increasing heliocentric distance. Grid lines on the ecliptic plane are 0.5 au apart. Cassini's path is shown south of the ecliptic, in Saturn's orbital plane tilted at 2.49$^{\circ}$, together with its projection on the ecliptic plane. As Cassini's motion was relatively slow compared to that of the comet, a wide range of solar wind velocities could have carried pickup protons from 153P to the spacecraft's vicinity. }
 \label{fig:fig2}
\end{figure}

In searching for an object that passed close to the Sun--Cassini line during this period, we investigated the trajectories of all comets listed in the IAU Catalogue of Cometary Orbits, IAU circulars, and minor planet circulars. During the period of interest, no planets or known comets were found to have approached as close to the Sun--Cassini line as comet 153P. Figure~\ref{fig:fig2} shows the geometry of the alignment between the comet and spacecraft. 153P crossed the ascending node of its orbit on 2002 March 9.02, DOY 68.02, at 0.556~au from the Sun, and reached perihelion north of the ecliptic on March 18.98, DOY 77.98. During the pickup proton enhancement period, taken as DOY 83 -- 105, March 24 -- April 15, Cassini's position, close to the comet's ascending node, was 9.78 million~km south of the ecliptic, and within 0.261 -- 0.296 au, 39 million~km, of the comet's orbital plane. Taking the solar wind's radial expansion into account, this corresponds to a minimum solar wind distance at the comet, when at a heliocentric distance of 0.556 au, of 3.02 -- 3.46 million km.

The sampling of a cometary ion tail downstream requires a combination of a spatial alignment between comet and spacecraft, and a solar wind velocity falling within the limited range required to carry the ions past the spacecraft at the time of alignment. Cassini's rate of motion during the period of enhanced pickup proton flux was relatively small, but the solar wind velocity still significantly affected the time of proton detection. 

Although the bulk solar wind plasma was not directly sampled during this period, the MIMI instrument  CHEMS sensor \citep{krimigis2004} did observe interstellar pickup ions, He$^{+}$. The cut-off in these ions' energy spectra can be used to derive solar wind speeds. The MIMI solar wind velocity (Figure~\ref{fig:fig1}(b)) was used to estimate the minimum approach distance to the comet of solar wind arriving at Cassini (Figure~\ref{fig:fig1}(d)). Perfectly radial solar wind flow of fixed speed was assumed. These assumptions were idealistic, but were the most reasonable assumptions to make in the absence of more information on the flow parameters and their histories. It should be borne in mind that the speed evolution of the solar wind as it traveled between comet and spacecraft, the possibility of non-radial flow, and the occurrence of faster or slower speeds than at shorter temporal scales than sampled could have affected the actual impact parameter, to both decrease or increase this minimum distance.
Assuming a radial flow, the solar wind with the minimum estimated impact parameter of $\sim$3.49 million km reached Cassini around DOY 90-95 (April 1-5), and had passed closest to the comet around DOY 66 (March 7).

As the comet's position was south of the ecliptic when passing closest to the Sun-Cassini line, the spacecraft probably traversed the northern side of the proton tail, i.e. northward of the main ion tail. Although there was clearly a large distance between Cassini and the comet's orbital plane at the time of the H$^{+}$ enhancement, it must be borne in mind that the originating neutral hydrogen corona was extremely extended, and that due to radial expansion, the proton tail's lateral extent was significantly greater than it was immediately downstream of the comet's corona. The extent of the neutral H corona is demonstrated by Lyman-$\alpha$ observations by the SWAN instrument aboard the SOHO spacecraft \citep{bertaux1995}. The radius of the neutral H cloud detectable by SWAN was approximately 0.1 au, or 15 million km, shown by the dashed line in Figure~\ref{fig:fig1}(d), corresponding to a lateral tail width of $\sim$2.36 au ($\sim$372 million km) at Cassini's heliocentric distance. The estimated 3.49 million km closest approach to the comet of wind sampled by Cassini was well within this extended neutral H corona.

\section{Solar wind feature correlations}

The Sun was at an elevated, post-solar maximum activity level during the period of interest. From remote observations of the Sun, and in-situ observations near Earth, it is clear that ambient solar wind conditions at the comet and Cassini were then highly variable, partly due to the presence of a coronal hole straddling the solar equator. In addition, the eruption of coronal mass ejecta, CMEs, was common. 153P is indeed known to have interacted with CMEs' interplanetary counterparts, ICMEs, during this time \citep{jones2004}.

The exact nature of the features seen in the magnetic field data is difficult to gauge in the absence of other solar wind plasma parameters at comparable temporal resolution. However, it is reasonable to assume that at least some of the magnetic field magnitude enhancements observed at Cassini are due to ICMEs. 
In comparing the features presented in Figure~\ref{fig:fig1}, it is noted that the initial increase in pickup proton detection coincides with a peak in $|$B$|$, and an increase in solar wind speed, v$_{sw}$. The direction of the magnetic field, not plotted here, underwent a fairly smooth rotation during DOY 87, suggesting that at least part of the structure was a magnetic cloud. If this feature is an ICME, it may suggest that a fast ejection swept up slower pickup protons already on their way to Cassini. An alternative explanation for a coincidence in ICME arrival and pickup proton flux increase is that a dense, fast ICME could significantly increase the rate of neutral hydrogen ionization in the comet's corona, i.e. the production rate of protons. The slight decrease in the apparent value of Q(H$_{2}$O) based on SOHO SWAN observations \citep{combi2008} around DOY 70, March 11, suggests that this latter explanation may be correct. As previously reported \citep{jones2004}, an ICME observed by the LASCO instrument on SOHO arrived at the comet on that day. 

Another, possibly more likely scenario is that the comet encountered non-radial flows, oriented a few degrees from the antisunward direction in the days around DOY 66. Flow deviations of a few degrees are very common, and it should be noted that cometary pickup ions from C/1999 T1 (McNaught-Hartley) were detected at the Ulysses spacecraft $\sim$12.4$^{\circ}$ from the Sun-comet line \citep{gloeckler2004}. Imposing this extreme angular separation as an upper limit to the detection of ions from 153P, the comet was conceivably within the range of detection by Cassini during DOY 62-70, i.e. March 3-11. This agrees with the independently-computed minimum in impact parameter shown in Figure~\ref{fig:fig1}(d). 

Two solar wind events during this DOY 62-70 period, both of which documented by \citet{jones2004}, could have diverted ions towards Cassini. On March 3 (DOY 62), 153P was further south of the ecliptic than was Cassini, at 0.62~au from the Sun, when it encountered an ICME. The local propagation direction of the cometary plasma, inferred from Earth-based images, was clearly northwards, i.e. towards Cassini's heliolatitude. 
The arrival of the ICME at the comet was observed as a major disruption of its ion tail, and was linked to a CME observed by the SOHO LASCO coronagraph erupting at 15:06 UT on March 2, with a plane-of-sky velocity of 1131 kms$^{-1}$ (Jones \& Brandt 2004). The minimum propagation velocity implied by the timing of the tail disruption was 912 kms$^{-1}$. Extrapolating this CME's perceived motion from 0.62 au to 7.1 au, the distance of Cassini, suggests an arrival at the spacecraft around DOY 74, March 15. If this event is caused by the ICME later detected by the magnetometer arriving at Cassini on March 16, when the spacecraft was 6.471 au downstream of the comet, a velocity of 862 kms$^{-1}$ is implied. The implied speeds of these two ICMEs are clearly higher than v$_{sw}$ derived from MIMI data. However, the latter data are averages over relatively long sampling periods, and do not resolve short-lived v$_{sw}$ variations.

On March 11-12 (DOY 70-71), by then north of the ecliptic, 153P appears to have encountered two ICMEs; the first driving ions southwards towards Cassini, the second in the opposite direction. It was suggested \citep{gloeckler2004} that pickup ions from C/1999 T1 were ducted towards Ulysses along magnetic field lines inside the CME. Such a magnetic connection with the comet is also a possible cause of the H$^{+}$ enhancements.

An aspect of these pickup proton enhancement events that must be addressed is the possibility that they were a result of compression of the solar wind alone. Such enhancements have been reported previously at high-latitude compression regions observed by Ulysses \citep{schwadron1999}. These were explained using an ion transport model that included compression but required a long scattering mean free path. We believe that this process is very unlikely to be the main cause of the prolonged H$^{+}$ enhancement at Cassini because of the lack of a He$^{+}$ enhancement coincident with the H$^{+}$, which differs from the Ulysses events. Also, at Ulysses, the compression regions recurred at the spacecraft at the solar rotation rate, whereas no such periodic H$^{+}$ enhancements were detected at Cassini, despite the quasi-periodic occurrence of $|$B$|$ enhancements at the spacecraft that were probably associated with the low-latitude coronal hole that existed at the time. It also should be noted that there are many localized increases in magnetic field magnitude that are not accompanied by an increase in pickup proton detection. 

\section{Conclusions}

Cometary pickup protons have previously been detected in-situ during cometary flyby missions, e.g. \citet{neugebauer1987}. Although the scale of the distance between comet 153P and Cassini is remarkable, the detection of cometary ions so far from their point of origin should not be unexpected. The detection of an enhanced He$^{+}$ population in the solar wind at 12$\times$10$^{6}$~km (0.08~au) from the nucleus of 1P/Halley was previously reported \citep{mihalov1987}; this population was interpreted as resulting from charge-exchange between solar wind $\alpha$ particles and cometary neutrals. Pickup ions originating at C/1999 T1 (McNaught-Hartley) were detected at Ulysses in 2000 after traveling a minimum distance of $\sim$1.05 au \citep{gloeckler2004}; the actual distance being difficult to measure as the event occurred within a CME. The ion tail of comet C/2006 P1 (McNaught) was traversed by the same spacecraft in 2007, $\sim$1.7 au downstream of the nucleus \citep{neugebauer2007}, and the tail of C/1996 B2 (Hyakutake) traversed in 1996 \citep{jones2000,gloeckler2000} after traveling $\sim$3.39 au antisunward, yielding an estimated tail length of $\sim$3.8 au. 
 
The radial distance traveled by the protons from comet 153P to Cassini was $\sim$6.56 au. During this time, the comet itself had however moved around its orbit. The range of possible tail lengths connecting the coma to the spacecraft was 7.46 -- 9.73 au, implying a minimum length greater than a billion km.

\section{Acknowledgements}

We dedicate this work to the memory of co-author Edward J. Smith, who made significant contributions to this work, but who sadly passed away in August 2019. Cassini-Huygens was a mission of international collaboration between NASA, the European Space Agency (ESA), and the Agenzia Spaziale Italiana (ASI). GHJ is grateful to the UK Science and Technology Facilities Council for partial support through consolidated grants ST/K000977/1 and ST/N000722/1, and to the Europlanet 2020 Research Infrastructure which received funding from the European Union's Horizon 2020 research and innovation programme under grant agreement No 654208. We are grateful for the use of data from the Cassini Magnetometer (PI: M. K. Dougherty, Imperial College London) and Magnetospheric Imaging Instrument, MIMI (PI: D. G. Mitchell, Johns Hopkins University Applied Physics Laboratory), and to J. T. T. M\"akinen of the Finnish Meteorological Institute, Helsinki, for guidance with SWAN data in the early stages of this project. The data used are listed in the references and the NASA Planetary Data System repository at https://pds.nasa.gov .

\bibliographystyle{elsarticle-num-names}
\bibliography{ikeyazhang.bib}

\begin{thebibliography}{14}
\expandafter\ifx\csname natexlab\endcsname\relax\def\natexlab#1{#1}\fi
\providecommand{\url}[1]{\texttt{#1}}
\providecommand{\href}[2]{#2}
\providecommand{\path}[1]{#1}
\providecommand{\DOIprefix}{doi:}
\providecommand{\ArXivprefix}{arXiv:}
\providecommand{\URLprefix}{URL: }
\providecommand{\Pubmedprefix}{pmid:}
\providecommand{\doi}[1]{\href{http://dx.doi.org/#1}{\path{#1}}}
\providecommand{\Pubmed}[1]{\href{pmid:#1}{\path{#1}}}
\providecommand{\bibinfo}[2]{#2}
\ifx\xfnm\relax \def\xfnm[#1]{\unskip,\space#1}\fi
\bibitem[{{Combi} et~al.(2008){Combi}, {M{\"a}kinen}, {Henry}, {Bertaux}, and
  {Quem{\'e}rais}}]{combi2008}
\bibinfo{author}{M.~R. {Combi}}, \bibinfo{author}{J.~T.~T. {M{\"a}kinen}},
  \bibinfo{author}{N.~J. {Henry}}, \bibinfo{author}{J.~L. {Bertaux}},
  \bibinfo{author}{E.~{Quem{\'e}rais}},
\newblock \bibinfo{title}{{Solar and Heliospheric Observatory/solar Wind
  Anisotropies Observations of Five Moderately Bright Comets: 1999-2002}},
\newblock \bibinfo{journal}{\aj} \bibinfo{volume}{135} (\bibinfo{year}{2008})
  \bibinfo{pages}{1533--1550}. \DOIprefix\doi{10.1088/0004-6256/135/4/1533}.
\bibitem[{{Young} et~al.(2004){Young}, {Berthelier}, {Blanc}, {Burch},
  {Coates}, {Goldstein}, {Grand e}, {Hill}, {Johnson}, {Kelha}, {McComas},
  {Sittler}, {Svenes}, {Szeg{\"o}}, {Tanskanen}, {Ahola}, {Anderson}, {Bakshi},
  {Baragiola}, {Barraclough}, {Black}, {Bolton}, {Booker}, {Bowman}, {Casey},
  {Crary}, {Delapp}, {Dirks}, {Eaker}, {Funsten}, {Furman}, {Gosling},
  {Hannula}, {Holmlund}, {Huomo}, {Illiano}, {Jensen}, {Johnson}, {Linder},
  {Luntama}, {Maurice}, {McCabe}, {Mursula}, {Narheim}, {Nordholt}, {Preece},
  {Rudzki}, {Ruitberg}, {Smith}, {Szalai}, {Thomsen}, {Viherkanto}, {Vilppola},
  {Vollmer}, {Wahl}, {W{\"u}est}, {Ylikorpi}, and {Zinsmeyer}}]{young2004}
\bibinfo{author}{D.~T. {Young}}, \bibinfo{author}{J.~J. {Berthelier}},
  \bibinfo{author}{M.~{Blanc}}, \bibinfo{author}{J.~L. {Burch}},
  \bibinfo{author}{A.~J. {Coates}}, \bibinfo{author}{R.~{Goldstein}},
  \bibinfo{author}{M.~{Grand e}}, \bibinfo{author}{T.~W. {Hill}},
  \bibinfo{author}{R.~E. {Johnson}}, \bibinfo{author}{V.~{Kelha}},
  \bibinfo{author}{D.~J. {McComas}}, \bibinfo{author}{E.~C. {Sittler}},
  \bibinfo{author}{K.~R. {Svenes}}, \bibinfo{author}{K.~{Szeg{\"o}}},
  \bibinfo{author}{P.~{Tanskanen}}, \bibinfo{author}{K.~{Ahola}},
  \bibinfo{author}{D.~{Anderson}}, \bibinfo{author}{S.~{Bakshi}},
  \bibinfo{author}{R.~A. {Baragiola}}, \bibinfo{author}{B.~L. {Barraclough}},
  \bibinfo{author}{R.~K. {Black}}, \bibinfo{author}{S.~{Bolton}},
  \bibinfo{author}{T.~{Booker}}, \bibinfo{author}{R.~{Bowman}},
  \bibinfo{author}{P.~{Casey}}, \bibinfo{author}{F.~J. {Crary}},
  \bibinfo{author}{D.~{Delapp}}, \bibinfo{author}{G.~{Dirks}},
  \bibinfo{author}{N.~{Eaker}}, \bibinfo{author}{H.~{Funsten}},
  \bibinfo{author}{J.~D. {Furman}}, \bibinfo{author}{J.~T. {Gosling}},
  \bibinfo{author}{H.~{Hannula}}, \bibinfo{author}{C.~{Holmlund}},
  \bibinfo{author}{H.~{Huomo}}, \bibinfo{author}{J.~M. {Illiano}},
  \bibinfo{author}{P.~{Jensen}}, \bibinfo{author}{M.~A. {Johnson}},
  \bibinfo{author}{D.~R. {Linder}}, \bibinfo{author}{T.~{Luntama}},
  \bibinfo{author}{S.~{Maurice}}, \bibinfo{author}{K.~P. {McCabe}},
  \bibinfo{author}{K.~{Mursula}}, \bibinfo{author}{B.~T. {Narheim}},
  \bibinfo{author}{J.~E. {Nordholt}}, \bibinfo{author}{A.~{Preece}},
  \bibinfo{author}{J.~{Rudzki}}, \bibinfo{author}{A.~{Ruitberg}},
  \bibinfo{author}{K.~{Smith}}, \bibinfo{author}{S.~{Szalai}},
  \bibinfo{author}{M.~F. {Thomsen}}, \bibinfo{author}{K.~{Viherkanto}},
  \bibinfo{author}{J.~{Vilppola}}, \bibinfo{author}{T.~{Vollmer}},
  \bibinfo{author}{T.~E. {Wahl}}, \bibinfo{author}{M.~{W{\"u}est}},
  \bibinfo{author}{T.~{Ylikorpi}}, \bibinfo{author}{C.~{Zinsmeyer}},
\newblock \bibinfo{title}{{Cassini Plasma Spectrometer Investigation}},
\newblock \bibinfo{journal}{\ssr} \bibinfo{volume}{114} (\bibinfo{year}{2004})
  \bibinfo{pages}{1--112}. \DOIprefix\doi{10.1007/s11214-004-1406-4}.
\bibitem[{{McComas} et~al.(2004){McComas}, {Schwadron}, {Crary}, {Elliott},
  {Young}, {Gosling}, {Thomsen}, {Sittler}, {Berthelier}, {Szego}, and
  {Coates}}]{mccomas2004}
\bibinfo{author}{D.~J. {McComas}}, \bibinfo{author}{N.~A. {Schwadron}},
  \bibinfo{author}{F.~J. {Crary}}, \bibinfo{author}{H.~A. {Elliott}},
  \bibinfo{author}{D.~T. {Young}}, \bibinfo{author}{J.~T. {Gosling}},
  \bibinfo{author}{M.~F. {Thomsen}}, \bibinfo{author}{E.~{Sittler}},
  \bibinfo{author}{J.~J. {Berthelier}}, \bibinfo{author}{K.~{Szego}},
  \bibinfo{author}{A.~J. {Coates}},
\newblock \bibinfo{title}{{The interstellar hydrogen shadow: Observations of
  interstellar pickup ions beyond Jupiter}},
\newblock \bibinfo{journal}{Journal of Geophysical Research (Space Physics)}
  \bibinfo{volume}{109} (\bibinfo{year}{2004}) \bibinfo{pages}{A02104}.
  \DOIprefix\doi{10.1029/2003JA010217}.
\bibitem[{{Dougherty} et~al.(2004){Dougherty}, {Kellock}, {Southwood},
  {Balogh}, {Smith}, {Tsurutani}, {Gerlach}, {Glassmeier}, {Gleim}, {Russell},
  {Erdos}, {Neubauer}, and {Cowley}}]{dougherty2004}
\bibinfo{author}{M.~K. {Dougherty}}, \bibinfo{author}{S.~{Kellock}},
  \bibinfo{author}{D.~J. {Southwood}}, \bibinfo{author}{A.~{Balogh}},
  \bibinfo{author}{E.~J. {Smith}}, \bibinfo{author}{B.~T. {Tsurutani}},
  \bibinfo{author}{B.~{Gerlach}}, \bibinfo{author}{K.~H. {Glassmeier}},
  \bibinfo{author}{F.~{Gleim}}, \bibinfo{author}{C.~T. {Russell}},
  \bibinfo{author}{G.~{Erdos}}, \bibinfo{author}{F.~M. {Neubauer}},
  \bibinfo{author}{S.~W.~H. {Cowley}},
\newblock \bibinfo{title}{{The Cassini Magnetic Field Investigation}},
\newblock \bibinfo{journal}{\ssr} \bibinfo{volume}{114} (\bibinfo{year}{2004})
  \bibinfo{pages}{331--383}. \DOIprefix\doi{10.1007/s11214-004-1432-2}.
\bibitem[{{Krimigis} et~al.(2004){Krimigis}, {Mitchell}, {Hamilton}, {Livi},
  {Dandouras}, {Jaskulek}, {Armstrong}, {Boldt}, {Cheng}, {Gloeckler}, {Hayes},
  {Hsieh}, {Ip}, {Keath}, {Kirsch}, {Krupp}, {Lanzerotti}, {Lundgren}, {Mauk},
  {McEntire}, {Roelof}, {Schlemm}, {Tossman}, {Wilken}, and
  {Williams}}]{krimigis2004}
\bibinfo{author}{S.~M. {Krimigis}}, \bibinfo{author}{D.~G. {Mitchell}},
  \bibinfo{author}{D.~C. {Hamilton}}, \bibinfo{author}{S.~{Livi}},
  \bibinfo{author}{J.~{Dandouras}}, \bibinfo{author}{S.~{Jaskulek}},
  \bibinfo{author}{T.~P. {Armstrong}}, \bibinfo{author}{J.~D. {Boldt}},
  \bibinfo{author}{A.~F. {Cheng}}, \bibinfo{author}{G.~{Gloeckler}},
  \bibinfo{author}{J.~R. {Hayes}}, \bibinfo{author}{K.~C. {Hsieh}},
  \bibinfo{author}{W.~H. {Ip}}, \bibinfo{author}{E.~P. {Keath}},
  \bibinfo{author}{E.~{Kirsch}}, \bibinfo{author}{N.~{Krupp}},
  \bibinfo{author}{L.~J. {Lanzerotti}}, \bibinfo{author}{R.~{Lundgren}},
  \bibinfo{author}{B.~H. {Mauk}}, \bibinfo{author}{R.~W. {McEntire}},
  \bibinfo{author}{E.~C. {Roelof}}, \bibinfo{author}{C.~E. {Schlemm}},
  \bibinfo{author}{B.~E. {Tossman}}, \bibinfo{author}{B.~{Wilken}},
  \bibinfo{author}{D.~J. {Williams}},
\newblock \bibinfo{title}{{Magnetosphere Imaging Instrument (MIMI) on the
  Cassini Mission to Saturn/Titan}},
\newblock \bibinfo{journal}{\ssr} \bibinfo{volume}{114} (\bibinfo{year}{2004})
  \bibinfo{pages}{233--329}. \DOIprefix\doi{10.1007/s11214-004-1410-8}.
\bibitem[{{Bertaux} et~al.(1995){Bertaux}, {Kyr{\"o}l{\"a}}, {Qu{\'e}merais},
  {Pellinen}, {Lallement}, {Schmidt}, {Berth{\'e}}, {Dimarellis}, {Goutail},
  {Taulemesse}, {Bernard}, {Leppelmeier}, {Summanen}, {Hannula}, {Huomo},
  {Kehl{\"a}}, {Korpela}, {Lepp{\"a}l{\"a}}, {Str{\"o}mmer}, {Torsti},
  {Viherkanto}, {Hochedez}, {Chretiennot}, {Peyroux}, and
  {Holzer}}]{bertaux1995}
\bibinfo{author}{J.~L. {Bertaux}}, \bibinfo{author}{E.~{Kyr{\"o}l{\"a}}},
  \bibinfo{author}{E.~{Qu{\'e}merais}}, \bibinfo{author}{R.~{Pellinen}},
  \bibinfo{author}{R.~{Lallement}}, \bibinfo{author}{W.~{Schmidt}},
  \bibinfo{author}{M.~{Berth{\'e}}}, \bibinfo{author}{E.~{Dimarellis}},
  \bibinfo{author}{J.~P. {Goutail}}, \bibinfo{author}{C.~{Taulemesse}},
  \bibinfo{author}{C.~{Bernard}}, \bibinfo{author}{G.~{Leppelmeier}},
  \bibinfo{author}{T.~{Summanen}}, \bibinfo{author}{H.~{Hannula}},
  \bibinfo{author}{H.~{Huomo}}, \bibinfo{author}{V.~{Kehl{\"a}}},
  \bibinfo{author}{S.~{Korpela}}, \bibinfo{author}{K.~{Lepp{\"a}l{\"a}}},
  \bibinfo{author}{E.~{Str{\"o}mmer}}, \bibinfo{author}{J.~{Torsti}},
  \bibinfo{author}{K.~{Viherkanto}}, \bibinfo{author}{J.~F. {Hochedez}},
  \bibinfo{author}{G.~{Chretiennot}}, \bibinfo{author}{R.~{Peyroux}},
  \bibinfo{author}{T.~{Holzer}},
\newblock \bibinfo{title}{{SWAN: A Study of Solar Wind Anisotropies on SOHO
  with Lyman Alpha Sky Mapping}},
\newblock \bibinfo{journal}{\solphys} \bibinfo{volume}{162}
  (\bibinfo{year}{1995}) \bibinfo{pages}{403--439}.
  \DOIprefix\doi{10.1007/BF00733435}.
\bibitem[{{Jones} and {Brandt}(2004)}]{jones2004}
\bibinfo{author}{G.~H. {Jones}}, \bibinfo{author}{J.~C. {Brandt}},
\newblock \bibinfo{title}{{The interaction of comet 153P/Ikeya-Zhang with
  interplanetary coronal mass ejections: Identification of fast ICME
  signatures}},
\newblock \bibinfo{journal}{\grl} \bibinfo{volume}{31} (\bibinfo{year}{2004})
  \bibinfo{pages}{L20805}. \DOIprefix\doi{10.1029/2004GL021166}.
\bibitem[{{Gloeckler} et~al.(2004){Gloeckler}, {Allegrini}, {Elliott},
  {McComas}, {Schwadron}, {Geiss}, {von Steiger}, and {Jones}}]{gloeckler2004}
\bibinfo{author}{G.~{Gloeckler}}, \bibinfo{author}{F.~{Allegrini}},
  \bibinfo{author}{H.~A. {Elliott}}, \bibinfo{author}{D.~J. {McComas}},
  \bibinfo{author}{N.~A. {Schwadron}}, \bibinfo{author}{J.~{Geiss}},
  \bibinfo{author}{R.~{von Steiger}}, \bibinfo{author}{G.~H. {Jones}},
\newblock \bibinfo{title}{{Cometary Ions Trapped in a Coronal Mass Ejection}},
\newblock \bibinfo{journal}{\apjl} \bibinfo{volume}{604} (\bibinfo{year}{2004})
  \bibinfo{pages}{L121--L124}. \DOIprefix\doi{10.1086/383524}.
\bibitem[{{Schwadron} et~al.(1999){Schwadron}, {Zurbuchen}, {Fisk}, and
  {Gloeckler}}]{schwadron1999}
\bibinfo{author}{N.~A. {Schwadron}}, \bibinfo{author}{T.~H. {Zurbuchen}},
  \bibinfo{author}{L.~A. {Fisk}}, \bibinfo{author}{G.~{Gloeckler}},
\newblock \bibinfo{title}{{Pronounced enhancements of pickup hydrogen and
  helium in high-latitude compressional regions}},
\newblock \bibinfo{journal}{\jgr} \bibinfo{volume}{104} (\bibinfo{year}{1999})
  \bibinfo{pages}{535--548}. \DOIprefix\doi{10.1029/1998JA900054}.
\bibitem[{{Neugebauer} et~al.(1987){Neugebauer}, {Goldstein}, {Goldstein},
  {Lazarus}, {Altwegg}, and {Balsiger}}]{neugebauer1987}
\bibinfo{author}{M.~{Neugebauer}}, \bibinfo{author}{B.~E. {Goldstein}},
  \bibinfo{author}{R.~{Goldstein}}, \bibinfo{author}{A.~J. {Lazarus}},
  \bibinfo{author}{K.~{Altwegg}}, \bibinfo{author}{H.~{Balsiger}},
\newblock \bibinfo{title}{{The pick-up of cometary protons by the solar wind}},
\newblock \bibinfo{journal}{\aap} \bibinfo{volume}{187} (\bibinfo{year}{1987})
  \bibinfo{pages}{21--24}.
\bibitem[{{Mihalov} et~al.(1987){Mihalov}, {Collard}, {Intriligator}, and
  {Barnes}}]{mihalov1987}
\bibinfo{author}{J.~D. {Mihalov}}, \bibinfo{author}{H.~R. {Collard}},
  \bibinfo{author}{D.~S. {Intriligator}}, \bibinfo{author}{A.~{Barnes}},
\newblock \bibinfo{title}{{Observation by Pioneer 7 of He $^{+}$ in the distant
  coma of Halley's comet}},
\newblock \bibinfo{journal}{\icarus} \bibinfo{volume}{71}
  (\bibinfo{year}{1987}) \bibinfo{pages}{192--197}.
  \DOIprefix\doi{10.1016/0019-1035(87)90172-2}.
\bibitem[{{Neugebauer} et~al.(2007){Neugebauer}, {Gloeckler}, {Gosling},
  {Rees}, {Skoug}, {Goldstein}, {Armstrong}, {Combi}, {M{\"a}kinen}, {McComas},
  {von Steiger}, {Zurbuchen}, {Smith}, {Geiss}, and
  {Lanzerotti}}]{neugebauer2007}
\bibinfo{author}{M.~{Neugebauer}}, \bibinfo{author}{G.~{Gloeckler}},
  \bibinfo{author}{J.~T. {Gosling}}, \bibinfo{author}{A.~{Rees}},
  \bibinfo{author}{R.~{Skoug}}, \bibinfo{author}{B.~E. {Goldstein}},
  \bibinfo{author}{T.~P. {Armstrong}}, \bibinfo{author}{M.~R. {Combi}},
  \bibinfo{author}{T.~{M{\"a}kinen}}, \bibinfo{author}{D.~J. {McComas}},
  \bibinfo{author}{R.~{von Steiger}}, \bibinfo{author}{T.~H. {Zurbuchen}},
  \bibinfo{author}{E.~J. {Smith}}, \bibinfo{author}{J.~{Geiss}},
  \bibinfo{author}{L.~J. {Lanzerotti}},
\newblock \bibinfo{title}{{Encounter of the Ulysses Spacecraft with the Ion
  Tail of Comet MCNaught}},
\newblock \bibinfo{journal}{\apj} \bibinfo{volume}{667} (\bibinfo{year}{2007})
  \bibinfo{pages}{1262--1266}. \DOIprefix\doi{10.1086/521019}.
\bibitem[{{Jones} et~al.(2000){Jones}, {Balogh}, and {Horbury}}]{jones2000}
\bibinfo{author}{G.~H. {Jones}}, \bibinfo{author}{A.~{Balogh}},
  \bibinfo{author}{T.~S. {Horbury}},
\newblock \bibinfo{title}{{Identification of comet Hyakutake's extremely long
  ion tail from magnetic field signatures}},
\newblock \bibinfo{journal}{\nat} \bibinfo{volume}{404} (\bibinfo{year}{2000})
  \bibinfo{pages}{574--576}.
\bibitem[{{Gloeckler} et~al.(2000){Gloeckler}, {Geiss}, {Schwadron}, {Fisk},
  {Zurbuchen}, {Ipavich}, {von Steiger}, {Balsiger}, and
  {Wilken}}]{gloeckler2000}
\bibinfo{author}{G.~{Gloeckler}}, \bibinfo{author}{J.~{Geiss}},
  \bibinfo{author}{N.~A. {Schwadron}}, \bibinfo{author}{L.~A. {Fisk}},
  \bibinfo{author}{T.~H. {Zurbuchen}}, \bibinfo{author}{F.~M. {Ipavich}},
  \bibinfo{author}{R.~{von Steiger}}, \bibinfo{author}{H.~{Balsiger}},
  \bibinfo{author}{B.~{Wilken}},
\newblock \bibinfo{title}{{Interception of comet Hyakutake's ion tail at a
  distance of 500 million kilometres}},
\newblock \bibinfo{journal}{\nat} \bibinfo{volume}{404} (\bibinfo{year}{2000})
  \bibinfo{pages}{576--578}. \DOIprefix\doi{10.1038/35007015}.

\end{thebibliography}







\end{document}